\documentclass[hyper,letterpaper,12pt]{article}
\usepackage{a4wide}
\usepackage{amsmath}
\usepackage[
      colorlinks=true,
      linkcolor=blue,
      urlcolor=blue,
      filecolor=black,
      citecolor=blue,
            ]{hyperref}
\usepackage{color}
\usepackage{graphicx}
\usepackage{amsfonts}
\usepackage{amssymb}
\usepackage{epstopdf}

\def\beq{\begin{equation}}
\def\eeq{\end{equation}}

\begin{document}
\title{\bf \Large Competition and Coexistence of Order Parameters in Holographic Multi-Band Superconductors}

\author{\large
~Rong-Gen Cai$^1$\footnote{E-mail: cairg@itp.ac.cn}~,
~~Li Li$^1$\footnote{E-mail: liliphy@itp.ac.cn}~,
~~Li-Fang Li$^1$$^2$\footnote{E-mail: lilf@itp.ac.cn}~,
~~Yong-Qiang Wang$^3$\footnote{E-mail: yqwang@lzu.edu.cn}\\
\\
\small $^1$State Key Laboratory of Theoretical Physics,\\
\small Institute of Theoretical Physics, Chinese Academy of Sciences,\\
\small Beijing 100190,  China.\\
\small $^2$State Key Laboratory of Space Weather, \\
\small Center for Space Science and Applied Research, Chinese Academy of Sciences,\\
\small Beijing 100190, China.\\
\small $^3$Institute of Theoretical Physics, Lanzhou University, Lanzhou, 730000, China.}
\date{\today}
\maketitle

\begin{abstract}
\normalsize We construct a holographic multi-band superconductor model with each complex scalar field in the bulk minimally coupled to a same gauge field. Taking into account the back reaction of matter fields on the background geometry and focusing on the two band case with two scalar order parameters,  we find that depending on the strength of the back reaction and the charge ratio of the two bulk scalars, five different superconducting phases exist, and three of five phases exhibit some region where both orders coexist and are thermodynamically favored. The other two superconducting phases have only one scalar order. The model exhibits rich phase structure and we construct the full diagram for the five superconducting phases. Our analysis indicates that the equivalent attractive interaction mediated by gravity between the two order parameters tends to make the coexistence of two orders much more easy rather than more difficult.
\end{abstract}

\tableofcontents

\section{ Introduction}

The AdS/CFT correspondence~\cite{Maldacena:1997re,Gubser:1998bc,Witten:1998qj} provides us a useful tool to study strongly coupled systems holographically in a dual framework through an appropriate gravity theory living in a higher dimensional space-time. One of the most studied objects is the so-called holographic superconductor. The first holographic model was constructed in refs.~\cite{Hartnoll:2008vx,Hartnoll:2008kx}. Since the condensed field is a scalar field dual to a scalar operator in the field theory side, it is a s-wave model. Such holographic approach was also generalized to the p-wave case~\cite{Gubser:2008wv} and d-wave case~\cite{Chen:2010mk,Benini:2010pr}. The basic idea is as follows.  Some matter fields, such as gauge fields and/or scalar fields, are added into the bulk. One of them  considered as  ``hair"  of the background solution plays the role of order parameter in the dual boundary system. As one tunes some parameters, such as temperature and chemical potential, the background solution without hair will become unstable and new stable solution with ``hair"  appears. This process in the bulk mimics the superconducting phase transition in the condensed matter theory.

Such simple holographic setups indeed uncover some basic properties of the real high $T_c$ superconducting materials. Nevertheless, most of studies existing in the literature have been based on a specific setup  where the dynamics in the bulk involves only one order parameter. It is desirable to generalize the single order parameter case to multi order parameter case.

The high $T_c$ superconducting systems, which are thought to be controlled by strongly coupled interactions, indeed involve various orders, such as magnetic ordering and superconductivity, see for example,  ref.~\cite{2009arXiv0901.4826B}. The holographic correspondence provides  us a convenient way to investigate the interaction for these orders by simply introducing dual fields in the bulk as well as appropriate couplings among them. The authors of ref.~\cite{Basu:2010fa} studied the competition between different orders in the holographic approach. They described a superconducting order by a charged scalar field and a magnetic order by a neutral scalar. It was showed that the appearance of one order inhibits the other one. But when the interactions between the bulk fields are repulsive enough  both orders can coexist and even enhance each other. The authors of ref.~\cite{Musso:2013ija} focused on the unbalanced holographic model by introducing two scalars charged under two U(1) gauge fields respectively. One accounts for the electrically superconducting order and the other indicates the electrically neutral magnetization. It was shown that the competition and enhancement between the two orders correspond to the attractive and repulsive interaction between the two scalar fields. In addition, the competition between superconducting and spatially modulated phases was exhibited in ref.~\cite{Donos:2012yu}.

On the other hand, some new high $T_c$ materials such as Magnesium diboride ($MgB_2$) and the iron pnictides ($LaFeAsO_{1-x}F_x$,$LiFeAs$, $Fe_{1+x}Se$, etc.) are characterized by multi Fermi surfaces and theoretical investigations have been based on multi-band models initiated from refs.~\cite{Kondo,Suhl,Leggett}. Such multi-band case can be phenomenologically described by the Ginzburg-Landau (GL) theory with several complex scalar fields minimally coupled to one gauge field. Within this framework, some significant properties of multi-band superconductors have been revealed, such as fractional flux vortices and vortex bound states (see, for example, refs.~\cite{2002PhRvL..89f7001B,2005PhRvB..71u4509S,2012JPSJ...81b4712Y}). One can build a holographic superconductor model involving more than one order parameters by straightforwardly generalizing the GL theory to the gravity side. Concretely,  one can consider a gravity theory with a negative cosmological constant, a gauge field as well as some complex scalar fields. The latter's condensates trigger the superconducting phase transition. In this setup, all scalar fields in the bulk are charged under a same U(1) gauge field.  Of course, there also exist other ways constructing holographic superconductor models with more than one order parameter. Treating three bulk scalars as the fundamental representation of SO(3) group, the authors of ref.~\cite{Huang:2011ac} studied a multi-band superconductor model, while regarding two bulk scalar fields as the fundamental representation of U(2) group, the authors of ref.~\cite{Krikun:2012yj} investigated a $S^{+/-}$ two-band superconductor model, where the AC conductivity exhibits a mid-infrared peak which is argued to be related to the inter-band interaction in iron-based superconductors.

The holographic model involving two competing scalar fields coupled to a single gauge field was first discussed in ref.~\cite{Basu:2010fa}, where the bulk scalar fields carry  different charges and masses, but have no direct interaction between them. The authors of ref.~\cite{Basu:2010fa} worked in the probe limit. Namely, they neglected the back reaction of matter fields on the background geometry, the planar Schwarzschild-AdS black hole. It was found that once a scalar field condenses, it will hinder the condensation  of the other one.~\footnote{The similar phenomenon was also noticed in the case of a non-Abelian model with SU(2) gauge symmetry in ref.~\cite{Basu:2008bh}.} Depending on model parameters, there are two different cases. The first case is that once a scalar condenses, the second one will never condense. The second case is more interesting. One scalar field first condenses, there the superconducting transition happens. When one lowers temperature,  the second scalar begins to condense, resulting in a phase where the both scalar fields condense. When one further lowers temperature, the first condensation quickly goes to zero at a certain temperature and the second condensation goes to a finite value. In the superconducting phases,  there exist three different cases: first scalar condenses only, the second scalar condenses only and both condense. One has to determine which case is thermodynamically favored. This can be done by comparing the free energy for each case. In addition, the probe limit is expected to be valid only near the critical point.~\footnote{Those studies~\cite{Basu:2010fa,Musso:2013ija,Krikun:2012yj,Huang:2011ac} we mentioned above are carried out in the probe approximation, neglecting the effect of matter fields on the background geometry.} In order to reach a more persuasive conclusion, one needs to consider the back reaction of matter fields on the bulk geometry. It was argued in ref.~\cite{Basu:2010fa} that once the back reaction of matter fields is taken into account, the coexistence region of the two orders will shrink. However, our study below shows that the back reaction will reveal more rich phase structure of the model. In addition, let us mention here that in a similar model with same mass and charge for two scalar fields, it was found~\cite{quasi} that in the probe limit, there is an unexpected gapless excitation with quadratic dispersion in the quasinormal
 mode spectrum.

We will construct a holographic multi-band superconductor model involving some scalar hairs in $(3+1)$ dimensional anti-de Sitter space-time in a rather succinct way by generalizing multi-band GL theories to the gravity side. The back reaction of matter fields will be considered.  As a concrete example, we will consider the two band case. The direct coupling between the two scalar fields, in some sense, can be effectively accountable with a simple shift of the effective mass of bulk scalars, which would not change the system very much~\cite{Musso:2013ija}. However, the situation is much more complicated by including the back reaction, since the effect of gravitational interaction can not be simply considered as only a shift of the effective mass for each bulk scalar. Therefore, we will ignore the direct coupling between two scalar fields. Of course, one of motivations of  this simple setup is to compare with the results in \cite{Basu:2010fa} and to see clearly the effect of the back reaction of matter fields. Here it is worth pointing out that the two scalars have an effective attractive interaction through gravity and this attractive interaction becomes stronger and stronger as one increases the strength of the back reaction.

In our study,  we can indeed find the coexistence region of two order parameters similar to the one studied in the probe limit~\cite{Basu:2010fa}. Nevertheless, we also find two additional new phases in which the two order parameters always coexist once they appear inside the superconducting phase.  As we increase the strength of the back reaction, our numerical calculation reveals that, for suitable parameters, the region of coexistence phases will enlarge, rather than shrink,  and finally the two order parameters always coexist as one increases the back reaction. For both new phases,  one of the scalar orders first condenses inducing the superconducting phase transition. Then the other order emerges at a lower temperature, which triggers a new phase transition within the superconducting phase. The similar new phase was also observed in a top down setting in ref.~\cite{Donos:2011ut}. The calculations of the conductivity and free energy uncover that the phase with two order parameters coexisting is indeed a superconducting phase and is thermodynamically favored. Adopting an eigenvalue method similar to the one in ref.~\cite{Horowitz:2013jaa}, we construct the parameter space for the coexistence phases. The system exhibits rich phase structure. Apart from the normal phase, we have totally five different superconducting phases. It is clear that including the back reaction is important to complete the full phases of the two band model. The full phase diagram for the five superconducting phases is constructed in figure~\ref{diagram}. Depending on the model parameters, i.e., the strength of the back reaction and the charge ratio between the two bulk scalars, each phase is the most thermodynamically stable phase in some region of the parameter space.

The paper is organized as follows. In section~\ref{sect:model}, we introduce the holographic model, obtain the equations of motion of the system and specify the boundary conditions to be satisfied. In section~\ref{sect:therPT}, we discuss the details of the phase transitions and study thermodynamic properties of the system. We calculate the optical conductivity in section~\ref{sect:conduc} to make sure that the new phases indeed describe superconducting state. We study, in section~\ref{sect:space}, the parameter space from which one can know the existence range of the coexistence phases, and construct the full phase diagram for all superconducting phases. The conclusion and further discussions are included in section~\ref{sect:conclusion}.


\section{The Holographic Model}
\label{sect:model}

In this paper, we will study a holographic superconductor model with $N$ scalar hairs in $(3+1)$ dimensional anti-de Sitter space-time. The action reads
\begin{equation}\label{action0}
S =\frac{1}{2\kappa^2}\int d^4 x
\sqrt{-g}[\mathcal{R}+\frac{6}{L^2}-\frac{1}{4}F_{\mu\nu} F^{\mu \nu}+\sum_{k=1}^{N}(-|\nabla\psi_k-ie_kA\psi_k|^2-m_k^2|\psi_k|^2)-\mathcal{V}_{intact}],
\end{equation}
where $L$ is the AdS radius, $\kappa^2\equiv 8\pi G $ is related to the gravitational constant in the bulk. $e_k$ and $m_k$ ($k=1,2,...,N$) are the charge and mass of the scalar field $\psi_k$, respectively. The term $\mathcal{V}_{intact}$ denotes the possible interaction among  bulk matter fields. $F_{\mu\nu}=\nabla_\mu A_\nu-\nabla_\nu A_\mu$ is field strength for the U(1) potential $A_\mu$.

We can see that the system depends on the mass $m_k$ and charge $e_k$ of each scalar field $\psi_k$. However, one can perform a rescaling of the type $A_\mu\rightarrow\frac{1}{e_2}A_\mu, \psi_k\rightarrow\frac{1}{e_2}\psi_k$ to set the charge of the scalar field $\psi_2$ to unity. Although we have $N$ scalar fields corresponding to $N$ order parameters in the dual boundary theory, the superconducting phase appears as long as one of the scalars has a nontrivial configuration. We are interested in the dynamics and mutual interaction among different orders. As a concrete example, we limit ourselves to the case with $N=2$.
The model we will study in this paper is described by the following action
\begin{equation}\label{action}
\begin{split}
S =\frac{1}{2\kappa^2}\int d^4 x
\sqrt{-g}[\mathcal{R}+\frac{6}{L^2}+\frac{1}{e_2^2}\mathcal{L}_m],\\
\mathcal{L}_m=-\frac{1}{4}F_{\mu\nu} F^{\mu \nu}-|D1\psi_1|^2-m_1^2|\psi_1|^2-|D2\psi_2|^2-m_2^2|\psi_2|^2,
\end{split}
\end{equation}
where we have defined $D1_\mu=\nabla_\mu-i\frac{e_1}{e_2}A_\mu$ and $D2_\mu=\nabla_\mu-iA_\mu$. The parameter $e_2$ controls the strength of the back reaction and $e_1/e_2$ is the effective charge of $\psi_1$ or the ratio of two scalar charges.

The equations of motion for the action~\eqref{action} read
\begin{equation}\label{eoms}
\begin{split}
D1^\mu D1_\mu\psi_1-m_1^2\psi_1=0, \\
D2^\mu D2_\mu\psi_1-m_2^2\psi_2=0, \\
\nabla^\mu F_{\mu\nu}=i\frac{e_1}{e_2}[\psi_1^*D1_\mu\psi_1-\psi_1 D1_\mu^*\psi_1^*]+i[\psi_2^*D2_\mu\psi_2-\psi_2 D2_\mu^*\psi_2^*],\\
\mathcal{R}_{\mu\nu}-\frac{1}{2}\mathcal{R}g_{\mu\nu}-\frac{3}{L^2}g_{\mu\nu}=\frac{1}{2e_2^2}F_{\mu\lambda}{F_\nu}^\lambda+\frac{1}{2e_2^2}[(D1_\mu\psi_1 D1_\nu^*\psi_1^*+D2_\mu\psi_2 D2_\nu^*\psi_2^*)+\mu\leftrightarrow\nu]+\frac{g_{\mu\nu}}{2e_2^2}\mathcal{L}_m.
\end{split}
\end{equation}
We would like to find static hairy black hole solutions to these equations, which mimic the superconducting phase.

\subsection{The ansatz and equations of motion}
\label{sect:ansatz}

The hairy black hole solution is assumed to take the following metric form
\begin{equation}\label{metric}
ds^2=-f(r)e^{-\chi(r)}dt^2+\frac{dr^2}{f(r)}+r^2(dx^2+dy^2),
\end{equation}
together with homogeneous  matter fields
\begin{equation}\label{matter}
\psi_1=\psi_1(r),\quad \psi_2=\psi_2(r),\quad A=\phi(r)dt.
\end{equation}
The horizon $r_h$ is determined by $f(r_h)=0$ and the temperature of the black hole is given by
\begin{equation}\label{temp}
T=\frac{f'(r_h)e^{-\chi(r_h)/2}}{4\pi}.
\end{equation}
One can use the U(1) gauge symmetry to set $\psi_1$ to be real. After using the $r$ component of Maxwell's equations we can also safely choose $\psi_2$ to be real.
We will work in the unites where $L=1$. Then, the independent equations of motion in terms of the above ansatz are deduced as follows
\begin{equation}\label{eoms}
\begin{split}
\psi_1''+(\frac{f'}{f}-\frac{\chi'}{2}+\frac{2}{r})\psi_1'+(\frac{e_1^2}{e_2^2}\frac{\phi^2e^{\chi}}{f^2}-\frac{m_1^2}{f})\psi_1=0, \\
\psi_2''+(\frac{f'}{f}-\frac{\chi'}{2}+\frac{2}{r})\psi_2'+(\frac{\phi^2e^{\chi}}{f^2}-\frac{m_2^2}{f})\psi_2=0, \\
\phi''+(\frac{\chi'}{2}+\frac{2}{r})\phi'-\frac{2}{f}(\frac{e_1^2}{e_2^2}\psi_1^2+\psi_2^2)\phi=0,\\
\frac{f'}{f}+\frac{r}{2e_2^2}(\psi_1'^2+\psi_2'^2)+\frac{re^{\chi}\phi'^2}{4e_2^2f}+\frac{r}{2e_2^2f}(m_1^2\psi_1^2+m_2^2\psi_2^2)+\frac{re^{\chi}\phi^2}{2e_2^2f^2}
(\frac{e_1^2}{e_2^2}\psi_1^2+\psi_2^2)-\frac{3r}{f}+\frac{1}{r}=0 ,\\
\chi'+\frac{r}{e_2^2}(\psi_1'^2+\psi_2'^2)+\frac{re^{\chi}\phi^2}{e_2^2f^2}(\frac{e_1^2}{e_2^2}\psi_1^2+\psi_2^2)=0,
\end{split}
\end{equation}
where a prime denotes the derivative with respect to $r$.

\subsection{Boundary conditions}
\label{sect:ansatz}
In order to find the solutions for all the five functions $\mathcal{F}=\{\psi_1,\psi_2,\phi,f,\chi\}$ one must specify suitable boundary conditions at both AdS boundary $r\rightarrow\infty$ and at the horizon $r=r_h$. In addition to $f(r_h)=0$, one must require $\phi(r_h)=0$ in order for $g^{\mu\nu}A_\mu A_\nu$ to be finite at the horizon.

In order to match the asymptotical AdS boundary, the general falloff of the matter and metric fields near the boundary $r\rightarrow\infty$ should behave as
\begin{equation} \label{boundary}
\begin{split}
\phi&=\mu-\frac{\rho}{r}+\ldots,\quad \psi_1=\frac{{\psi_1}_-}{r^{{\Delta_1}_-}}+\frac{{\psi_1}_{+}}{r^{{\Delta_1}_+}}+\ldots, \quad \psi_2=\frac{{\psi_2}_-}{r^{{\Delta_2}_-}}+\frac{{\psi_2}_{+}}{r^{{\Delta_2}_+}}+\ldots, \\
\quad f&=r^2+\frac{\epsilon}{r}+\ldots,\quad \chi=0+\ldots,
\end{split}
\end{equation}
where ${\Delta_i}_\pm=\frac{3\pm\sqrt{9+4 {m_i}^2}}{2}$ ($i=1,2$). According to the AdS/CFT correspondence, we impose ${\psi_i}_-=0$, since we want the U(1) symmetry to be broken spontaneously.  Following the AdS/CFT dictionary, up to a normalization, the coefficients $\mu$, $\rho$ and ${\psi_i}_{+}$ are interpreted as chemical potential, charge density and the expectation value of the scalar operator $\mathcal{O}_i$ in the dual field theory, respectively.

Regularity of the solution at the horizon $r=r_h$ requires that all our functions have finite values and admit a Taylor series expansion in terms of $(r-r_h)$ as
\begin{equation}\label{series}
\mathcal{F}=\mathcal{F}(r_h)+\mathcal{F}'(r_h)(r-r_h)+\cdots.
\end{equation}
By plugging the expansion~\eqref{series} into~\eqref{eoms}, one can find that there are five independent parameters at the horizon $\{r_h,\psi_1(r_h),\psi_2(r_h),\phi'(r_h),\chi(r_h)\}$.
Note that the equations of motion~\eqref{eoms} have two scaling symmetries
\begin{equation} \label{scaling1}
e^{-\chi} \rightarrow \lambda^2 e^{-\chi},\quad \phi\rightarrow\lambda\phi,\quad t\rightarrow \lambda^{-1}t,
\end{equation}
\begin{equation} \label{scaling2}
r\rightarrow\lambda r,\quad (t,x,y)\rightarrow{\lambda^{-1}}(t,x,y),\quad f\rightarrow\lambda^2 f,\quad \phi\rightarrow\lambda \phi.
\end{equation}
Taking advantage of the two scaling symmetries, we can first set $\{r_h=1,\chi(r_h)=0\}$ for performing numerics. After solving the coupled differential equations, we should use the first symmetry again to satisfy the asymptotic condition $\chi(\infty)=0$. Thus we have three independent parameters $\{\psi_1(r_h),\psi_2(r_h),\phi'(r_h)\}$, where two of them will be chosen as shooting parameters to match the source free conditions, i.e, ${\psi_1}_-=0$ and ${\psi_2}_-=0$. After solving the set of equations, we can obtain the condensate $\langle\mathcal{O}_i\rangle$, chemical potential $\mu$ and charge density $\rho$ by reading off the corresponding coefficients in~\eqref{boundary}, respectively.\footnote{In our unites, the AdS/CFT dictionary gives $ \langle \mathcal{O}_i \rangle_{real}=\frac{2\Delta_{i+}-3}{\kappa^2 e_2^2}\psi_{i+}$ and $\rho_{real}=\frac{1}{2\kappa^2 e_2^2}\rho$. In what follows, we neglect these prefactors and this will not change our conclusions. }

The normal phase in the dual field theory is characterized by the vanishing vacuum expectation value of both condensates $\mathcal{O}_1$ and $\mathcal{O}_2$, which corresponds to vanishing scalar fields $\psi_1$ and $\psi_2$ in the bulk. The gravity background describing the normal phase is just the Reissner-Nordstr\"om-AdS black hole with planar symmetry
\begin{equation}\label{normal}
\phi(r)=\mu(1-\frac{r_h}{r}),\quad \psi_1(r)=\psi_2(r)=0,\quad f(r)=r^2(1-\frac{r_h^3}{r^3})+\frac{r_h^2}{4r^2}\frac{\mu^2}{e_2^2}(1-\frac{r}{r_h}),
\end{equation}
where $r_h$ is the black hole horizon and $\mu$ is the chemical potential of the black hole.
\section{Thermodynamics and Phase Transition}
\label{sect:therPT}
As we can see from~\eqref{action}, the two band model is controlled by four model parameters, i.e., $m_1^2$, $m_2^2$, $e_2$, and $e_1/e_2$. We will choose $m_1^2=0$ and $m_2^2=-2$ in this paper.~\footnote{One of the reasons for this choice is for the convenience in numerical calculations. The other is to compare our results with those in the probe limit~\cite{Basu:2010fa}.}  Following the study in \cite{Basu:2010fa}, one may expect that the model admits three different superconducting phases. The first superconducting phase corresponds to the case with $\psi_1\neq0$ and $\psi_2=0$ (Phase-\uppercase\expandafter{\romannumeral1}). The second superconducting phase corresponds to the case with $\psi_2\neq0$ and $\psi_1=0$ (Phase-\uppercase\expandafter{\romannumeral2}). The third superconducting phase admits the region where both of the scalars condense.

\begin{figure}[h]
\centering
\includegraphics[scale=0.92]{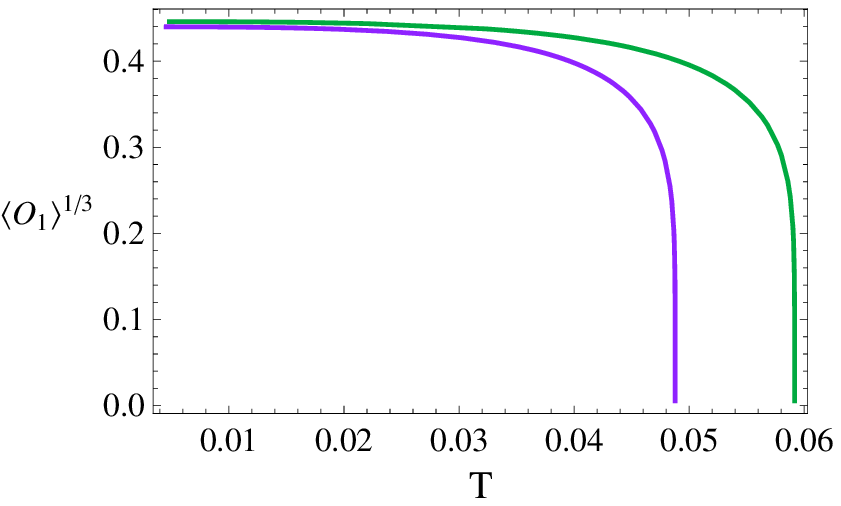}\ \ \ \
\includegraphics[scale=0.9]{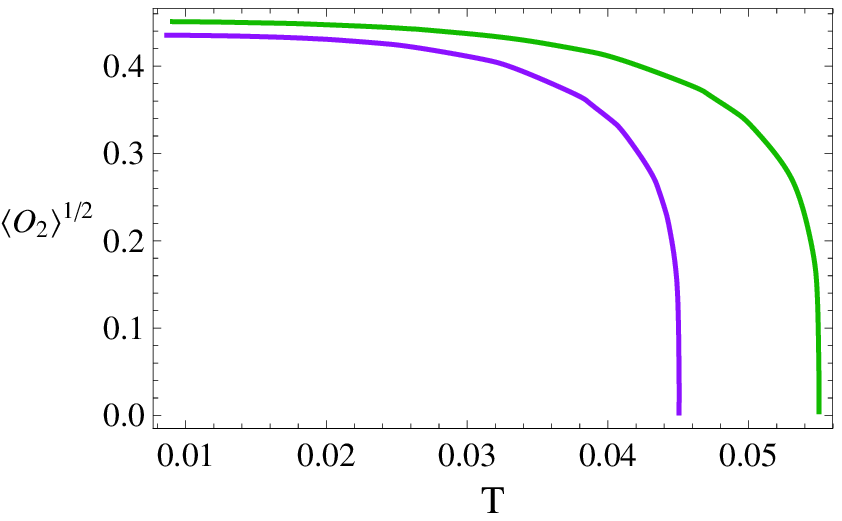} \caption{\label{condensate12} The condensate as a function of temperature for Phase-\uppercase\expandafter{\romannumeral1} with $m_1^2=0$ in the left plot and Phase-\uppercase\expandafter{\romannumeral2} with $m_2^2=-2$ in the right plot, respectively. The green curves correspond to $e_1/e_2=2$ and $e_2=4$, and the purple lines correspond to $e_1/e_2=2$ and $e_2=2$.}
\end{figure}

The first two cases with only one scalar condensing were discussed in ref.~\cite{Hartnoll:2008kx}. The condensate for a single scalar hair as a function of temperature with a fixed chemical potential for such two cases is drawn in figure~\ref{condensate12}. The condensate curves are very similar to each other for different parameters. As we lower  temperature, the normal phase becomes unstable to developing scalar hair at a certain critical temperature $T_c$.  The critical temperature $T_c$ will decrease if we increase the strength of the back reaction.
\begin{figure}[h]
\centering
\includegraphics[scale=1.2]{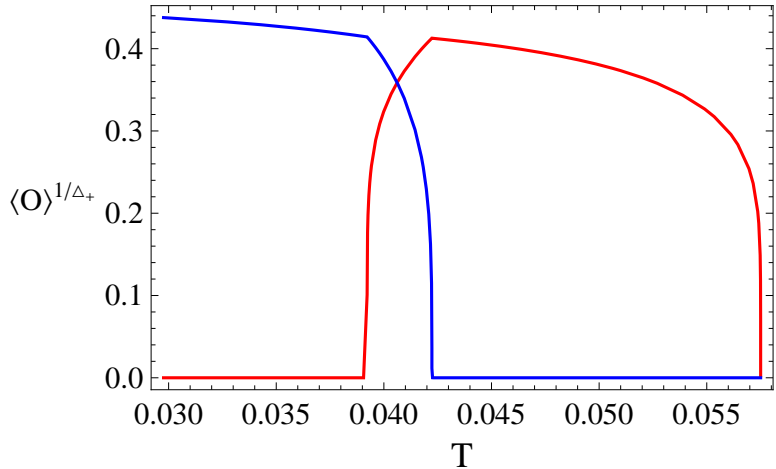} \caption{\label{condensateC} The condensate as a function of temperature for Phase-C. The red curve is for $\psi_1$, while the blue one is for $\psi_2$. We set $e_1/e_2=1.95$ and $e_2=4$.}
\end{figure}

 Except for these two cases with a single scalar condensate, we find other three kinds of condensate behaviors where the model admits the coexistence region of two scalar condensates. The first one is the case uncovered in the probe limit in ref.~\cite{Basu:2010fa}. This superconducting phase denoted by Phase-C is drawn in figure~\ref{condensateC}. One can see clearly from the figure that as we lower temperature, the scalar $\psi_1$ first condenses at $T_c$ where the superconducting phase transition happens; when we continue lowering temperature to a certain value, say $T_2$,  the scalar $\psi_2$ begins to condense, while the condensate of $\psi_1$ decreases, resulting in the phase with both orders; if one further lowers temperature, the first condensate quickly goes to zero at a temperature, say, $T_3$; when temperature is lower than $T_3$, there exists only the condensate of $\psi_2$.

\begin{figure}[h]
\centering
\includegraphics[scale=0.9]{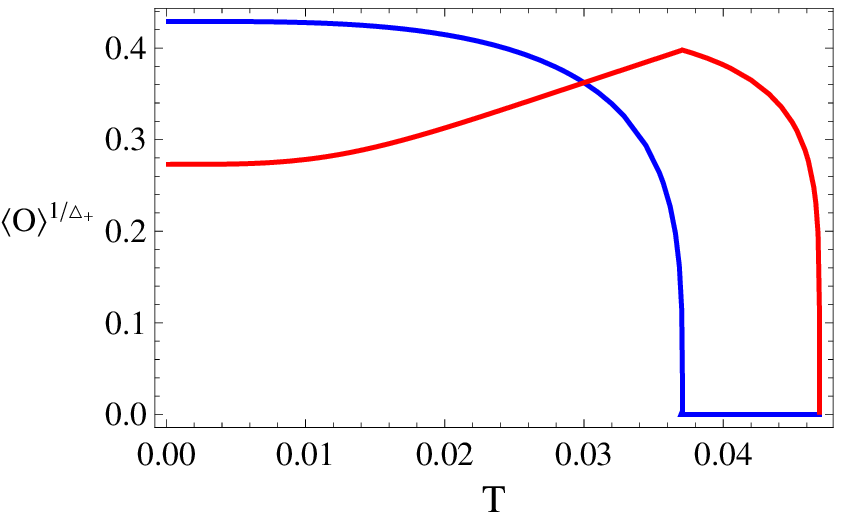}\ \ \ \
\includegraphics[scale=0.9]{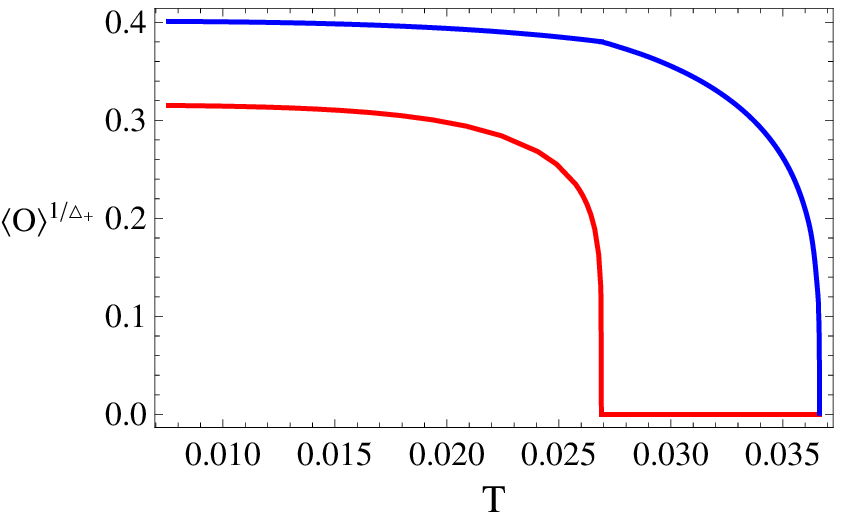} \caption{\label{condensateAB} The condensate as a function of temperature for Phase-A with $e_1/e_2=1.95$ and $e_2=2$ in the left plot and Phase-B with $e_1/e_2=1.9$ and $e_2=1.5$ in the right plot, respectively. The red curves correspond to $\psi_1$, and the blue curves correspond to $\psi_2$.}
\end{figure}

 The second case is presented in the left plot of figure~\ref{condensateAB}. We can see from the plot that the condensate of $\psi_1$ first happens at the critical temperature $T_c$; when one lowers temperature, the condensate of $\psi_2$ emerges at a temperature, say $T_2$; if one continues lowering temperature, the condensate of $\psi_2$ increases while $\psi_1$ decreases, but the latter will never go to zero.  In this case, when temperature is less than $T_2$,  both orders always coexist. We denote this case by Phase-A. Depending on the back reaction, the inverse is also true:  the condensate of $\psi_1$ emerges following the condensate of $\psi_2$, and then both orders are always present. This case is labeled as Phase-B drawn in the right plot of figure~\ref{condensateAB}.   Whether these three coexistence phases make physical sense depends on whether they are thermodynamically favored in their own parameter spaces.   Here let us mention that in our setup, for a given $e_2$, the critical temperature $T_c$ for Phase-B is fixed, while the critical temperature $T_c$ for Phase-A increases as one increases the value of $e_1/e_2$.

\subsection{Thermodynamics}
\label{sect:thermo}
 Thus we have totally five different superconducting phases in our model. In order to determine which phase is the thermodynamically favored phase in some parameter space,
 we should calculate free energy of the system for each phase. We will work in grand canonical ensemble in this paper, where the chemical potential is fixed. In gauge/gravity duality the grand potential $\Omega$ of the boundary thermal state is identified with temperature times the on-shell bulk action with Euclidean signature. The Euclidean action must include the Gibbons-Hawking boundary term for a well-defined Dirichlet variational principle and further a surface counter term for removing divergence
\begin{equation}
\begin{split}
2\kappa^2 S_{Euclidean}=-\int d^4 x
\sqrt{g}[&\mathcal{R}+\frac{6}{L^2}-\frac{1}{e_2^2}\mathcal{L}_m]+\int_{r\rightarrow\infty} d^3x
\sqrt{h}(-2\mathcal{K}+\frac{4}{L})\\
+&\frac{1}{e_2^2}\int_{r\rightarrow\infty} d^3x\sqrt{h}(\frac{{\triangle_1}_+-3}{L}\psi_1^2+\frac{{\triangle_2}_+-3}{L}\psi_2^2),
\end{split}
\end{equation}
where $h$ is the induced metric on the boundary $r\rightarrow\infty$, and $\mathcal{K}$ is the trace of the extrinsic
curvature. By using of the equations of motion~\eqref{eoms} and the asymptotical expansion of matter and metric functions near the AdS boundary, the grand potential $\Omega$ can be expressed as
\begin{equation}\label{grand1}
\frac{2\kappa^2\Omega}{V_2}=\epsilon,
\end{equation}
with $V_2=\int dxdy$. More specifically, for the normal phase shown in~\eqref{normal}, one has $\epsilon=-r_h^3-\frac{r_h}{4}\frac{\mu^2}{e_2^2}$.

\begin{figure}[h]
\centering
\includegraphics[scale=0.82]{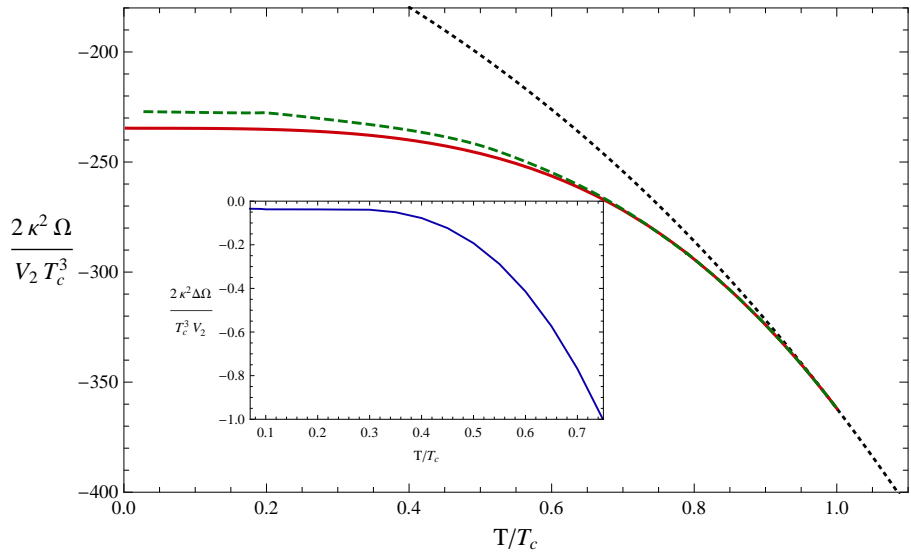} \caption{\label{freeA} The free energy as a function of temperature for Phase-A, denoted by solid red curve. We also plot other possible phases in the same figure. The black dotted line is from the normal phase and the green dashed one corresponds to Phase-\uppercase\expandafter{\romannumeral1}. The curve in the insert is the difference of free energy between Phase-A and Phase-\uppercase\expandafter{\romannumeral2}. We choose $e_1/e_2=2$ and $e_2=2$. The critical superconducting temperature where Phase-A appears in this case is $T_c\simeq0.0469\mu$, which is also the critical temperature of Phase-\uppercase\expandafter{\romannumeral1}. The temperature where $\psi_2$ begins to condense for Phase-A is about $0.0307\mu$, and the critical temperature of Phase-\uppercase\expandafter{\romannumeral2} is about $0.0451\mu$.}
\end{figure}

The free energy corresponding to phase-A is drawn in figure~\ref{freeA}. Since in this case, we has three additional solutions from our equations of motion~\eqref{eoms}, corresponding to the normal phase, Phase-\uppercase\expandafter{\romannumeral1} and Phase-\uppercase\expandafter{\romannumeral2}. In order to make sure  whether phase-A is the most thermodynamically stable phase, we also plot the free energy of these three phases in the same figure. Indeed, the three superconducting phases have lower free energy than the normal phase. Furthermore, Phase-A does have the lowest free energy among the three superconducting phases, indicating that once Phase-A appears, it is thermodynamically favored. Comparing Phase-B with Phase-A, the only difference is that $\psi_2$ condenses before $\psi_1$. Therefore, the free energy of Phase-B also dominates once it appears.

\begin{figure}[h]
\centering
\includegraphics[scale=0.9]{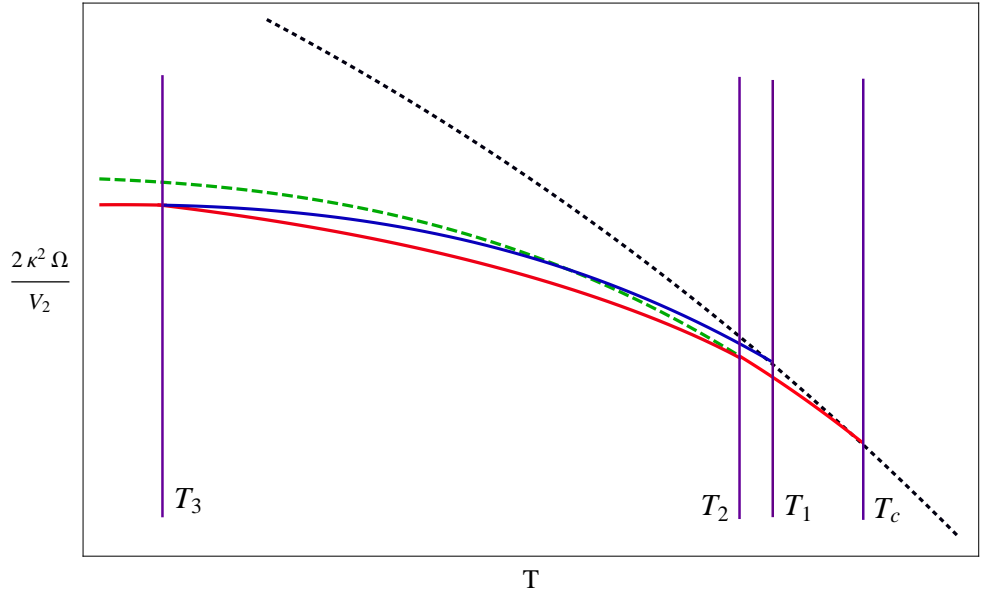} \caption{\label{freeC} The typical configuration for the free energy as a function of temperature for Phase-C labeled as the solid red curve. In this case, the equations of motion also admit other three types of solutions, i.e., the normal phase (dotted black curve), Phase-\uppercase\expandafter{\romannumeral1} (dashed green curve) and Phase-\uppercase\expandafter{\romannumeral2} (solid blue curve). There are four special temperature denoted as $T_c$, $T_1$, $T_2$ and $T_3$, corresponding to the critical temperature for the superconducting transition, the critical temperature for Phase-\uppercase\expandafter{\romannumeral2}, the temperature at which $\psi_2$ begins to appear in Phase-C, and the temperature where $\psi_1$  goes to zero in Phase-C, respectively. Phase-C and Phase-\uppercase\expandafter{\romannumeral1} coincide as $T_2<T<T_c$, so do Phase-C and Phase-\uppercase\expandafter{\romannumeral2} as $T<T_3$.}
\end{figure}

The free energy corresponding to Phase-C is schematically drawn in figure~\ref{freeC}. Similar to the above discussion, we also present free energy for other allowable solutions in the same figure. As we know, in Phase-C, there is only a narrow window admitting the two orders to coexist. Outside this region, it reduces to phases with only a single order. More specifically, at the beginning of the superconducting transition, there is no condensate of $\psi_2$, so Phase-C coincides with Phase-\uppercase\expandafter{\romannumeral1}. As $\psi_2$ begins to condense at a lower temperature, the free energy of Phase-C becomes lower than Phase-\uppercase\expandafter{\romannumeral1} as well as Phase-\uppercase\expandafter{\romannumeral2}. For a much lower temperature, $\psi_1$ vanishes and Phase-C coincides with Phase-\uppercase\expandafter{\romannumeral2} then. All in all, Phase-C is the most thermodynamically favored phase over other possible phases including the normal phase, Phase-\uppercase\expandafter{\romannumeral1} and Phase-\uppercase\expandafter{\romannumeral2}.

\subsection{Superconducting Phase Transition}
\label{sect:phase}
As we have seen, for suitable parameters, the solutions with coexisting $\psi_1$ and $\psi_2$ appear, including Phase-A, Phase-B and Phase-C. Furthermore, such phases are thermodynamically favored, compared to the superconducting phases with a single order parameter, i.e., Phase-\uppercase\expandafter{\romannumeral1} and Phase-\uppercase\expandafter{\romannumeral2}. We now give a further discussion for such three thermodynamically stable phases.

\begin{figure}[h]
\centering
\includegraphics[scale=1.1]{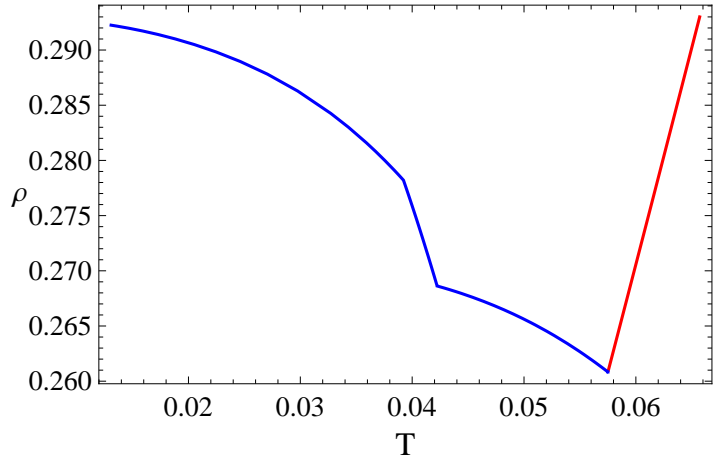} \caption{\label{chargeC} The charge density as a function of temperature for Phase-C. The red curve is for the normal phase, while the blue one corresponds to the superconducting phase. We set $e_1/e_2=1.95$ and $e_2=4$. There are three special temperatures at which the derivative of charge density with the temperature is discontinuous.}
\end{figure}

Phase-C is similar to the case found in ref.~\cite{Basu:2010fa}, where the emergency of the scalar $\psi_2$ suppresses the condensate of the scalar $\psi_1$ even although $\psi_1$ condenses before $\psi_2$. Phase-C exhibits three continuous phase transitions. It can be seen clearly from figure~\ref{chargeC}, where the charge density versus temperature is plotted,  there exist three particular points at which the derivative of the charge density with respect to temperature is discontinuous, indicating a second order transition. The one with the highest special temperature is the critical point for the superconducting phase transition, while the remaining two points inside the superconducting phase. We have checked that the entropy of the black hole solution as a function of temperature also exhibits the same discontinuity at these special points.

\begin{figure}[h]
\centering
\includegraphics[scale=1.1]{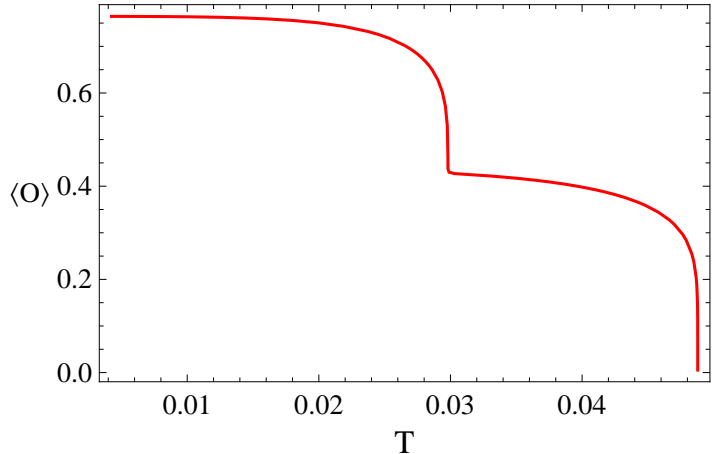} \caption{\label{condensateAA} The total condensate as a function of temperature for Phase-A. We set $e_1/e_2=2$ and $e_2=2$. The two special points at $T_c\simeq0.0488\mu$ and $T\simeq0.0298\mu$ correspond to the superconducting critical point and the point at which $\psi_2$ begins to emerge in Phase-A, respectively.}
\end{figure}
The condensate behaviors  and  other thermodynamical properties are similar for Phase-A and Phase-B. There exist two transition points in both cases. One is the critical superconducting phase transition and the other within the superconducting phase, which indicates the fact that our system is multi-band. In order to see this clearly, we define the total condensate as $\langle\mathcal{O}\rangle=\langle O_{1+}\rangle^{1/\Delta_{1+}}+\langle O_{2+}\rangle^{1/\Delta_{2+}}$, and draw $\langle\mathcal{O}\rangle$ as a function of temperature in figure~\ref{condensateAA}. As one lowers temperature, $\langle\mathcal{O}\rangle$ emerges at the critical superconducting phase transition point, then at a certain temperature inside the superconducting phase it has a sudden increase, where the condensate of the other $\psi$ appears. Such behavior is very reminiscent of the one in the real multi-band superconductor, see, for example, the figure~3 or figure~12 in ref.~\cite{2005PhRvB..71u4509S}. Our discussion can be straightforwardly  generalized to the case with $N\geq3$ order parameters. For suitable parameters, we believe that the total condensate as a function of temperature would exhibit as many as $N$ sudden increases, indicating appearance of a new order parameter there.

 It was argued in ref.~\cite{Basu:2010fa} that as one considers the back reaction, the coexistence region  for two order parameters in Phase-C will shrink. Surprisedly, comparing the two plots in figure~\ref{condensateC} and figure~\ref{condensateAB}, one can clearly see that if one increases the strength of the back reaction, the region of coexistence phases will enlarge, rather than shrink and the two order parameters finally always coexist as the back reaction is strong enough. Phase-A and Phase-B are new phases which are not found in the probe limit~\cite{Basu:2010fa}. In particular, the probe limit analysis uncovered that for the chosen masses, i.e., $m_1^2=0$ and $m_2^2=-2$, $\psi_1$ will never condense if $\psi_2$ condenses firstly. Nevertheless, our results show that it is not the case once the back reaction is taken into account. One can indeed find the Phase-B in the right plot of figure~\ref{condensateAB}, where $\psi_1$ condenses following $\psi_2$ and then both coexist.

 In some regions of parameter space, the equations of motion do not admit black hole solutions corresponding to Phase-A, Phase-B and Phase-C, instead what we can obtain is the case with only one scalar condensate, i.e., either Phase-\uppercase\expandafter{\romannumeral1} or Phase-\uppercase\expandafter{\romannumeral2}. For a given temperature, the thermodynamically favored phase is obtained by comparing the free energy of Phase-\uppercase\expandafter{\romannumeral1} and Phase-\uppercase\expandafter{\romannumeral2}. One might expect that either Phase-\uppercase\expandafter{\romannumeral1} or Phase-\uppercase\expandafter{\romannumeral2} is always thermodynamically favored over the other. However, there might  be an interesting possiblity that one of the two phases is thermodynamically favored at the beginning, and as we lower temperature, the other phase might become more thermodynamically stable. Thus, the physical state is the competition between Phase-\uppercase\expandafter{\romannumeral1} and Phase-\uppercase\expandafter{\romannumeral2}, and the charge density as well as the entropy will be discontinuous at a certain point inside the superconducting phase, indicating a first order phase transition. In our calculations we have not found such a possibility.

\section{Conductivity}
\label{sect:conduc}
\begin{figure}[h]
\centering
\includegraphics[scale=0.92]{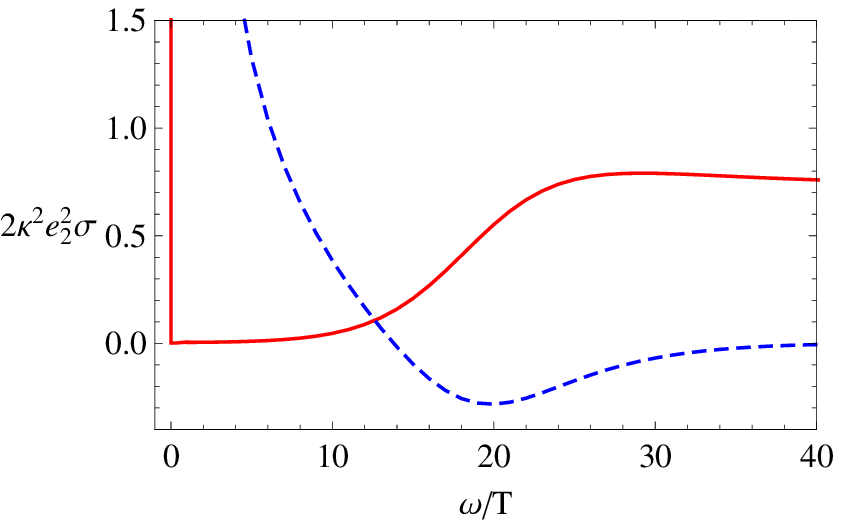}\ \ \ \
\includegraphics[scale=0.92]{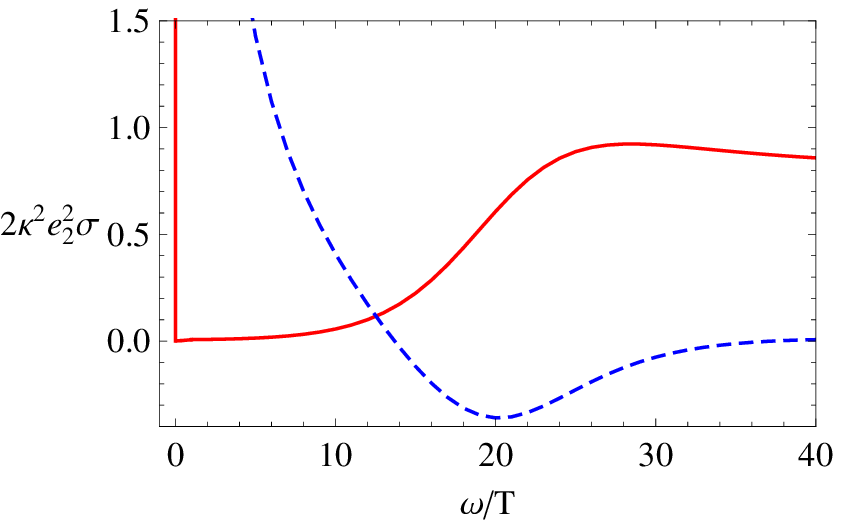} \caption{\label{conductivity} The optical conductivity as a function of frequency at temperature $T=0.0273\mu$ for Phase-A (left plot) and Phase-\uppercase\expandafter{\romannumeral1} (right plot), respectively. The red solid line is the real part of the conductivity, while the blue dashed line is the imaginary part of the conductivity. We choose the parameter $e_1/e_2=1.95$ and $e_2=2$ and the critical temperature of Phase-A is $T_c\simeq0.0469\mu$. There is a delta function at the origin for the real part of the conductivity in both cases.}
\end{figure}

We have found that there exist five superconducting phases in the holographic two-band model. In order to ensure the system is indeed in a superconducting state and to see whether there are any new phenomena occurring in such a two-band model, we would like to calculate the conductivity $\sigma$. Since we move away from the probe limit, we have to consider the fluctuations of $A_x$ and $g_{tx}$. Assuming both perturbations have a time dependence of the form $e^{-i\omega t}$, we can obtain the equations of motion for $A_x$ and $g_{tx}$ by linearizing the full equations of motion~\eqref{eoms}, which read
\begin{equation}\label{axeom}
A_x''+(\frac{f'}{f}-\frac{\chi'}{2})A_x'+[\frac{\omega^2}{f^2}e^{\chi}-\frac{2}{f}(\frac{e_1^2}{e_2^2}\psi_1^2+\psi_2^2)]A_x+\frac{\phi'}{f}e^{\chi}(g_{tx}'-\frac{2}{r}g_{tx})=0,
\end{equation}
and
\begin{equation}\label{gtxeom}
g_{tx}'-\frac{2}{r}g_{tx}+\frac{A_x}{e_2^2}\phi'=0.
\end{equation}
Substituting~\eqref{gtxeom} into~\eqref{axeom}, we obtain the final equation of motion to calculate the conductivity
\begin{equation}\label{Axeom}
A_x''+(\frac{f'}{f}-\frac{\chi'}{2})A_x'+[(\frac{\omega^2}{f^2}-\frac{\phi'^2}{e_2^2f})e^{\chi}-\frac{2}{f}(\frac{e_1^2}{e_2^2}\psi_1^2+\psi_2^2)]A_x=0.
\end{equation}
Since the conductivity is related to the retarded two-point function of the U(1) current, i.e, $\sigma=\frac{1}{i\omega}G^R(\omega,k=0)$, we impose the ingoing boundary condition near the horizon
\begin{equation}\label{ingoing}
A_x=(r-r_h)^{-\frac{i\omega}{4\pi T}}[a_0+a_1(r-r_h)+a_2(r-r_h)^2+\cdots],
\end{equation}
with $a_0, a_1, a_2$ being constants. The gauge field $A_x$ near the boundary $r\rightarrow\infty$ falls off as
\begin{equation}\label{axbound}
A_x=A^{(0)}+\frac{A^{(1)}}{r}+\cdots.
\end{equation}
According to the AdS/CFT dictionary, the retarded Green function can be read as $G^R=\frac{1}{2\kappa^2 e_2^2}\frac{A^{(1)}}{A^{(0)}}$, from which one can obtain the conductivity
\begin{equation}\label{conduc}
\sigma(\omega)=\frac{1}{i\omega}G^R(\omega,k=0)=\frac{1}{2\kappa^2 e_2^2}\frac{A^{(1)}}{i\omega A^{(0)}}.
\end{equation}
The optical conductivity as a function of frequency in the region with two order parameters is presented in the left plot of figure~\ref{conductivity}. For a comparison,  we also draw the conductivity for the case with only one order in the right plot of figure~\ref{conductivity}. We can see clearly that the optical conductivity in two band model behaves qualitatively similar to the  model with only one scalar order~\cite{Hartnoll:2008kx}. In addition, from the Kramers-Kronig relations, one can conclude that the real part of the conductivity has a Dirac delta function at $\omega=0$ since the imaginary part has a pole, i.e., Im$[\sigma(\omega)]\sim\frac{1}{\omega}$.

\section{Parameter Space and Phase Diagram}
\label{sect:space}
The calculations of the free energy and conductivity reveal that the phase with coexistence of two order parameters is indeed a superconducting phase and is thermodynamically stable. More specifically, depending on the model parameters, one gets two kinds of  condensate behaviors. One is the case that the order parameter emerging at lower temperature suppresses the condensate of the first scalar, the latter will go to zero quickly (Phase-C). The other is the case that once the two orders appear, they always coexist, including Phase-A and Phase-B.

It is helpful to construct the parameter space form which one can learn in which region the superconducting orders can coexist.  However, to numerically find the solution space where the two superconducting orders coexist is a little difficult. By our shooting method, it is more easy to find the single order phase rather than the phase with two orders. Nevertheless, one can complete this task  by just turning the problem around. Focus on the concrete model discussed in our paper, i.e., $m_1^2=0$ and $m_2^2=-2$, a good starting point is to find the critical valve of the ratio $e_1/e_2$ such that $T$ is a critical temperature at which $\psi_1$ begins to vanish or emerge. At such a temperature, $\psi_1$ is very small and can be treated as a perturbation on the background where only $\psi_2$ condenses
\begin{equation}\label{conduc}
-\psi_1''-(\frac{f'}{f}-\frac{\chi'}{2}+\frac{2}{r})\psi_1'+\frac{m_1^2}{f}\psi_1=\frac{e_1^2}{e_2^2}\frac{\phi^2e^{\chi}}{f^2}\psi_1,
\end{equation}
where $\{\phi, f, \chi\}$ are functions describing the hairy AdS black hole with only $\psi_2$ non-vanishing.

\begin{figure}[h]
\centering
\includegraphics[scale=0.9]{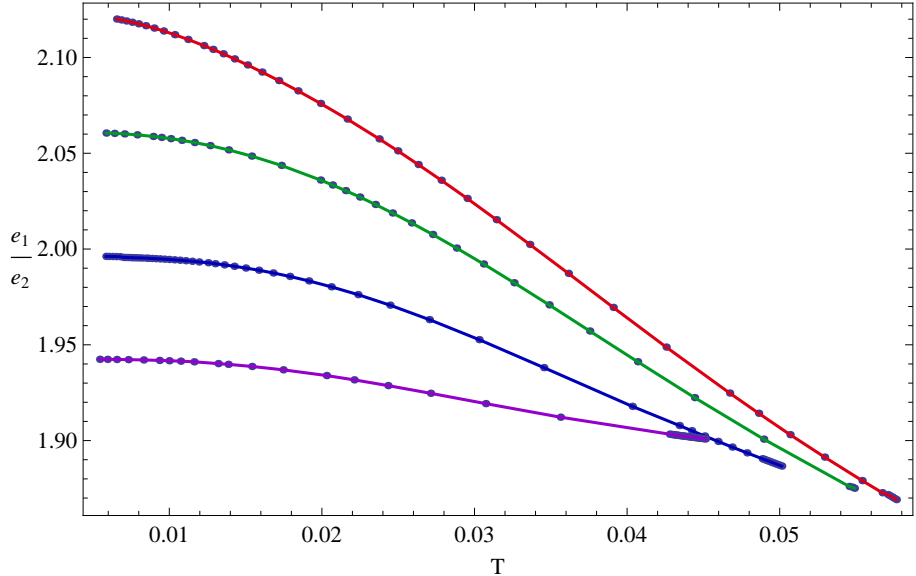}
\caption{\label{parameter1} The parameter space for small back reaction. For each $e_2$, we plot the ratio $e_1/e_2$ as a function of the critical temperature at which the solutions with a single $\psi_2$ become unstable to developing new scalar hair. Every point gives the value of $e_1/e_2$ and the corresponding temperature below which the order $\psi_1$ tends to vanish.  The curves from up to down correspond to $e_2=8.0, 4.0, 2.6, 2.0$, respectively.}
\end{figure}

\begin{figure}[h]
\centering
\includegraphics[scale=0.9]{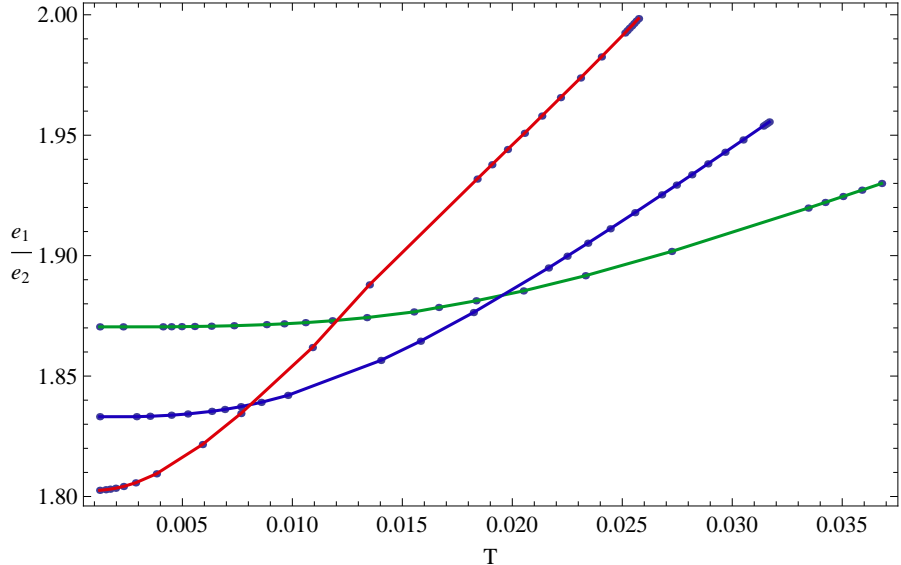}
\caption{\label{parameter2} The parameter space for large back reaction. For each $e_2$, we plot the ratio $e_1/e_2$ with respect to the critical temperature at which the solutions with a single $\psi_2$ become unstable to developing hair $\psi_1$. Each point gives the value of $e_1/e_2$ and the corresponding temperature below which the order $\psi_1$ begins to appear.  From the up to down in the leftmost the curves correspond to $e_2=1.5, 1.3, 1.1$, respectively.}
\end{figure}

We demand $\psi_1$ to be regular at the horizon and to fall off as in~\eqref{boundary} near the AdS boundary. Then this equation can be considered as an eigenvalue problem with positive eigenvalue $e_1^2/e_2^2$.~\footnote{A similar method has been adopted in ref.~\cite{Horowitz:2013jaa}.} The parameter space is presented in figure~\ref{parameter1} and figure~\ref{parameter2}, where different curves correspond to different strengths of back reaction $1/e_2^2$. Comparing the two figures, we can find that the ratio $e_1/e_2$ versus the critical temperature at which the solutions with a single scalar $\psi_2$ become unstable to developing new scalar hair behaves differently for large and small back reactions.

Figure~\ref{parameter1} corresponds to the case with large $e_2$, i.e., small back reaction. The critical ratio $e_1/e_2$ on the curve for a given $e_2$ decreases with respect to temperature, and there is a one-to-one correspondence between each point on the curve and Phase-C.  Concretely, in Phase-C, the temperature below which the order $\psi_1$ vanishes is just the temperature given here with the corresponding value of $e_1/e_2$ that can be read on the curve. One can further find that each curve has a maximum ${(e_1/e_2)}_{max}$ at $T \approx 0$ as well as a minimum  ${(e_1/e_2)}_{min}$ at $T=T_{max}$, which means that Phase-C can appear only when the ratio $e_1/e_2$ is in between them, i.e., $e_1/e_2\in[{(e_1/e_2)}_{min},{(e_1/e_2)}_{max}]$. For the case $e_1/e_2>{(e_1/e_2)}_{max}$, Phase-A appears and dominates the system. If we decrease the value $e_1/e_2$, $\psi_2$ will condense before $\psi_1$. Our eigenvalue method tells us that the superconducting background with $\psi_2$ hair only does not become unstable under the perturbation imposed by $\psi_1$, which means $\psi_1$ can never appear inside the condensate of $\psi_2$. This is just the case denoted as Phase-\uppercase\expandafter{\romannumeral2}. Nevertheless, the case with $\psi_1\neq0$ and $\psi_2=0$ labeled as Phase-\uppercase\expandafter{\romannumeral1} is also the possible phase. Thus, we have two simple phases with only a single order, i.e., Phase-\uppercase\expandafter{\romannumeral1} and Phase-\uppercase\expandafter{\romannumeral2} when $e_1/e_2<{(e_1/e_2)}_{min}$. For a given temperature, the thermodynamically favored state is the consequence of the competition between Phase-\uppercase\expandafter{\romannumeral1} and Phase-\uppercase\expandafter{\romannumeral2}. As we discussed earlier, one of the two phases would always be thermodynamically preferred to the other. And the other possibility is that one of the two phases is firstly thermodynamically favored, then the other phase becomes more thermodynamically favored, which indicates a first order phase transition. However, our numerical calculation rules out the later possibility and the thermodynamically favored state is Phase-\uppercase\expandafter{\romannumeral2}. As one can see, as  the strength of the back reaction increases, the value of ${(e_1/e_2)}_{min}$ and ${(e_1/e_2)}_{max}$ for each curve increases and decreases, respectively, which indicates that the attractive interaction mediated by gravity reduces the parameter space of Phase-C, but enlargers the parameter space of Phase-A.

Let us now move to the case with large back reaction, which is presented in figure~\ref{parameter2}. In contrary to the small back reaction case, for a given $e_2$, the critical ratio $e_1/e_2$ on the curve increases with respect to temperature and has a maximum  $(T_{max},{(e_1/e_2)}_{max})$ and a minimum $(T \approx 0,{(e_1/e_2)}_{min})$. In figure~\ref{parameter1}, the point located in the range $[{(e_1/e_2)}_{min},{(e_1/e_2)}_{max}]$ is related to Phase-C, while the point in figure~\ref{parameter2} here corresponds to Phase-B. For $e_1/e_2<{(e_1/e_2)}_{min}$ and $e_1/e_2>{(e_1/e_2)}_{max}$, one can obtain phases with only one order parameter. The physical state might be in Phase-\uppercase\expandafter{\romannumeral1}, Phase-\uppercase\expandafter{\romannumeral2} or the combination of both. Nevertheless, by comparing the free energy between Phase-\uppercase\expandafter{\romannumeral1} and Phase-\uppercase\expandafter{\romannumeral2}, we find that Phase-\uppercase\expandafter{\romannumeral1} is thermodynamically favored for $e_1/e_2>{(e_1/e_2)}_{max}$ and Phase-\uppercase\expandafter{\romannumeral2} is thermodynamically favored for $e_1/e_2<{(e_1/e_2)}_{min}$. We does not find any suitable values for $e_1/e_2$ and $e_2$ for which one can have a phase similar to Phase-C. In contrary to the small back reaction case, we find that, as one strengthens the back reaction, ${(e_1/e_2)}_{min}$ decreases and ${(e_1/e_2)}_{max}$ increases, respectively. Therefore, the attractive interaction between the two bulk scalar fields mediated by gravity enhances the parameter space of Phase-B. Comparing the two figures, it is clear that there must be a certain value of $e_2$ denoted by $e_2^{critical}$ at which the curve becomes parallel to horizontal axis. At $e_2^{critical}$, ${(e_1/e_2)}_{max}$ has the same value of ${(e_1/e_2)}_{min}$, thus there is no region for the Phase-C in figure~\ref{parameter1} as well as Phase-B in figure~\ref{parameter2} to survive. In our model discussed in this paper $e_2^{critical}\simeq1.762$.

\begin{figure}[h]
\centering
\includegraphics[scale=1.0]{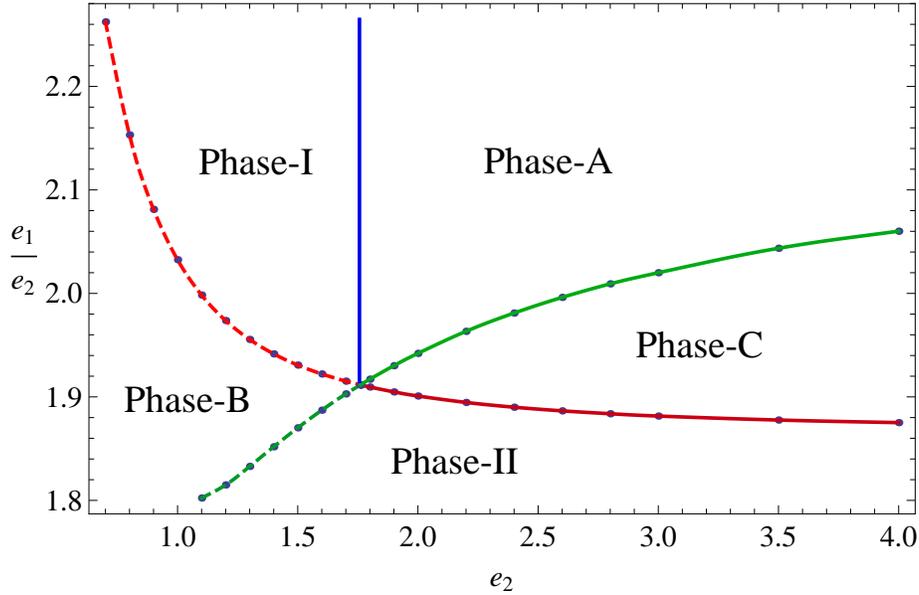}
\caption{\label{diagram} The full phase diagram for the five superconducting phases. Depending on $e_1/e_2$ and $e_2$, the phase diagram is divided into five parts. We label the most thermodynamically favored phase in each part.}
\end{figure}

To summarize, apart from the normal phase, we obtain as many as five superconducting phases in our model. Depending on the model parameters $e_1/e_2$ and $e_2$, each phase can be most thermodynamically stable in some region of parameter space. The full phase diagram for the five superconducting phases is constructed in figure~\ref{diagram} which is divided into five parts by five boundaries. The corresponding phase we named in each region is the most thermodynamically stable phase. The solid red line between Phase-C and Phase-\uppercase\expandafter{\romannumeral2} and the dashed red line between Phase-B and Phase-\uppercase\expandafter{\romannumeral1} is obtained by finding the value of $e_1/e_2$ at which the two critical temperatures for the phases with only $\psi_1$ or $\psi_2$ condensate are the same for each given $e_2$. The solid green curve separating Phase-C from Phase-A and the dashed green curve separating Phase-B from Phase-\uppercase\expandafter{\romannumeral2} correspond to the leftmost points in figure~\ref{parameter1} and figure~\ref{parameter2}. The vertical blue line dividing Phase-A from Phase-\uppercase\expandafter{\romannumeral1} is given by the critical line at $e_2^{critical}$. We see from figure~\ref{diagram} that when the value of $e_2$ increases, the upper bound of Phase-C (solid green curve) increases, but finally it arrives at a finite one around $2.164$. In contrast, the lower bound of Phase-C (solid red curve) decreases, and finally it arrives at a finite one around $1.867$.~\footnote{Due to the lake of the numerical control for very large back reaction, we can not fix the phase boundaries for very small $e_2$. However, our current results indicate that the boundary of Phase-B tends to expand as one increases the strength of the back reaction.}

As we increase the strength of the back reaction, the region for Phase-C with the coexisting behavior of two order parameters only in a narrow window is gradually forced to shrink and finally vanishes at $e_2^{critical}$, while the regions for Phase-A and Phase-B where both order parameters always present enlarge. In this sense, we can conclude that the gravity which provides an equivalent attractive interaction between the holographic order parameters tends to make the coexistence of two orders much more easy rather than more difficult. The other way to see this is to compare different curves for given ratio $e_1/e_2$. We can find from figure~\ref{parameter1} that increasing the strength of the back reaction lowers the temperature at which $\psi_1$ vanishes. Thus the coexistence region in Phase-C can survive at much lower temperature. On the other hand, in the case with not very small $e_1/e_2$ in figure~\ref{parameter2}, we can see that increasing the strength of the back reaction uplifts the temperature where $\psi_2$ background becomes unstable to forming $\psi_1$ hair, resulting in a much wider region where both order parameters can coexist.


\section{Conclusion and further discussions}
\label{sect:conclusion}

We have investigated a holographic superconductor model with more than one order parameter in four dimensions, where each complex scalar field in the bulk is minimally coupled to a same  U(1) gauge field. This can be interpreted as a holographic multi-band superconductor model. We have studied the interplay among different macroscopic orders in this model, where the back reaction of matters on the background geometry has been taken into account.

Concretely, we have discussed the two-band case with mass squares $m_1^2=0$ and $m_2^2=-2$ for two bulk scalar fields $\psi_1$ and $\psi_2$, respectively. Depending on the strength of the back reaction $1/e_2^2$ and the relative charge ratio $e_1/e_2$  of the two scalar fields, the model admits as many as five different superconducting phases. Three of them, denoted by Phase-A, Phase-B and Phase-C, exhibit the coexistence region of two order parameters. More specifically, for Phase-C, as we lower temperature, the second scalar $\psi_2$ condenses following $\psi_1$  will completely suppress the condensate of the first order, i.e., $\psi_1$ will go to zero finally. The condensate behaviors in Phase-A and Phase-B are similar. One of the two orders condenses first, and once the other begins to condense, both always coexist.  Other two superconducting phases, labeled as Phase-\uppercase\expandafter{\romannumeral1} and Phase-\uppercase\expandafter{\romannumeral2}, have only one order parameter. For details, see figure~\ref{condensate12}, figure~\ref{condensateC} and figure~\ref{condensateAB}, respectively.

For given parameters, i.e., $e_1/e_2$ and $e_2$, Phase-A, Phase-B and Phase-C can not appear at the same time, while Phase-\uppercase\expandafter{\romannumeral1} and Phase-\uppercase\expandafter{\romannumeral2} are always allowable. Therefore, once we have a phase with coexisting orders, we can also have the two phases which have only one order. We have calculated the free energy for each possible phase and found that the three phases with coexisting orders are thermodynamically favored than Phase-\uppercase\expandafter{\romannumeral1} and Phase-\uppercase\expandafter{\romannumeral2}. For each $e_2$, we plot the ratio $e_1/e_2$ as a function of the temperature at which the solutions with only $\psi_2$ condensate become unstable to developing new scalar hair $\psi_1$ in figure~\ref{parameter1} and figure~\ref{parameter2}. The behavior of $e_1/e_2$ versus temperature is exactly opposite for small and large back reaction cases. The model shows rich phases structure and we construct the full phase diagram in figure~\ref{diagram}. From such three figures, we can conclude that the gravity providing an equivalent attractive interaction between the holographic order parameters in some sense will enhance the coexistence region and tend to make the coexistence of two orders much more easy rather than more difficult. It also implies that to reveal a full phase structure for a holographic model, the back reaction of matter fields in the bulk has to be taken into account.

In this paper, we have considered in some detail the case with $m_1^2=0$ and $m_2^2=-2$. We expect that the whole picture will not be qualitatively changed  for other choices of mass parameters if one always regards the scalar with larger mass as $\psi_1$ and the scalar with smaller mass as $\psi_2$. A special case is that the two scalars have equal mass. In this case, both condensates can appear at the same temperature for some parameters, for example, $e_1/e_2=1$ even with no direct coupling between two scalar fields. One may worry that in the region which only admits the solutions corresponding to Phase-\uppercase\expandafter{\romannumeral1} or Phase-\uppercase\expandafter{\romannumeral2}, there might be a first order phase transition between Phase-\uppercase\expandafter{\romannumeral1} and Phase-\uppercase\expandafter{\romannumeral2} by comparing their free energy. We have checked a lot of parameter values and found that either Phase-\uppercase\expandafter{\romannumeral1} or Phase-\uppercase\expandafter{\romannumeral2} is always dominant. Nevertheless, even although such a first order transition might exist, the full phase diagram does not change. It is interesting to understand whether this first order phase transition could appear or not for other possible values of mass. In addition,  in our discussions, we have turned off the direct coupling between two bulk scalars, it is also interesting to study the case with a direct interaction between the two scalars. Of course, our study can also be straightforwardly generalized to the case with more than two order parameters. In that case we expect much richer phase structure will appear.

According to the symmetry of the macroscopic wave function or condensate of Cooper pairs in the real superconducting materials, the superconductor can be classified by s-wave, p-wave, d-wave and so on. The holographic s-wave, p-wave and d-wave superconductor models are already available in the literature, it is therefore quite interesting to study the holographic models with superconducting order parameters with different spins. Furthermore, we may study the holographic multi-band superconductor model by including the effect of lattice~\cite{Horowitz:2013jaa}. In our present study, the gravity background is chosen to be black holes, which corresponds to the superconductor/conductor phase transition at finite temperature. Note that the holographic superconductor/insulator phase transition at zero temperature  has been studied in ref.~\cite{Nishioka:2009zj} by choosing the so called AdS soliton background with one spatial direction compactified to a circle and the complete phase diagram for such a system was constructed in ref.~\cite{Horowitz:2010jq}. Certainly it is very interesting to investigate the holographic superconductor/insulator phase transition with more than one order parameter.  We leave these issues for further study.

\section*{Acknowledgements}

This work was supported in part by the National Natural Science Foundation of China (No.10821504, No.10975168,  No.11035008, No.11205226, No.11175019 and No.41231066), and
in part by the Ministry of Science and Technology of China under Grant No.2010CB833004. L.F.L would like to appreciate the National Basic Research Program of China (973 Program) Grant No.2011CB811404 and the Specialized Research Fund for State Key Laboratories. Y.Q.W was supported by the National Natural Science Foundation of China (No.11005054). L.L would like to thank Yan Liu and Ya-Wen Sun for useful discussions.

\appendix

\end{document}